\documentclass[sigconf]{acmart}
\usepackage{booktabs} 
\usepackage{float}
\usepackage{multirow}
\usepackage{graphicx}
\usepackage{xcolor}
\usepackage{soul}
\usepackage[utf8]{inputenc}
\usepackage{subcaption}
\usepackage{xfrac}
\usepackage{listings}
\newcommand{\ignore}[1]{}
\AtBeginDocument{%
  }

\copyrightyear{2026}
\acmYear{2026}
\setcopyright{cc}
\setcctype{by-nc-nd}
\acmConference[CHI '26]{Proceedings of the 2026 CHI Conference on Human Factors in Computing Systems}{April 13--17, 2026}{Barcelona, Spain}
\acmBooktitle{Proceedings of the 2026 CHI Conference on Human Factors in Computing Systems (CHI '26), April 13--17, 2026, Barcelona, Spain}
\acmPrice{}
\acmDOI{10.1145/3772318.3791473}
\acmISBN{979-8-4007-2278-3/2026/04}

\begin{document}

\title[Negotiating Digital Identities with AI Companions]{Negotiating Digital Identities with AI Companions: Motivations, Strategies, and Emotional Outcomes}

\author{Renkai Ma}
\authornote{Both authors contributed equally to this research and share first authorship.}
\orcid{0000-0002-4434-2235}
\email{mark@ucmail.uc.edu}
\affiliation{
  \department{School of Information Technology}
  \institution{University of Cincinnati}
  \city{Cincinnati}
  \state{Ohio}
  \country{USA}
}

\author{Shuo Niu}
\authornotemark[1]
\orcid{0000-0002-8316-4785}
\email{shniu@clarku.edu}
\affiliation{
  \department{Department of Computer Science}
  \institution{Clark University}
  \city{Worcester}
  \state{Massachusetts}
  \country{USA}
}

\author{Lingyao Li}
\email{lingyaol@usf.edu}
\orcid{0000-0001-5888-8311}
\affiliation{
  \department{School of Information}
  \institution{University of South Florida}
  \city{Tampa}
  \state{Florida}
  \country{USA}
}

\author{Alex Hirth}
\email{ahirth@clarku.edu}
\orcid{0009-0006-8403-7143}
\affiliation{
  \institution{Clark University}
  \city{Worcester}
  \state{Massachusetts}
  \country{USA}
}

\author{Ava Brehm}
\email{ABrehm@clarku.edu}
\orcid{0009-0006-5728-4982}
\affiliation{
  \institution{Clark University}
  \city{Worcester}
  \state{Massachusetts}
  \country{USA}
}

\author{Rowajana Behterin Barbie}
\email{rbarbie@clarku.edu}
\orcid{0009-0003-6303-3643}
\affiliation{
  \institution{Clark University}
  \city{Worcester}
  \state{Massachusetts}
  \country{USA}
}

\renewcommand{\shortauthors}{Ma et al.}

\begin{abstract}
  AI companions enable deep emotional relationships by engaging a user's sense of identity, but they also pose risks like unhealthy emotional dependence. Mitigating these risks requires first understanding the underlying process of identity construction and negotiation with AI companions. Focusing on Character.AI (C.AI), a popular AI companion, we conducted an LLM-assisted thematic analysis of 22,374 online discussions on its subreddit. Using Identity Negotiation Theory as an analytical lens, we identified a three-stage process: 1) five user motivations; 2) an identity negotiation process involving three communication expectations and four identity co-construction strategies; and 3) three emotional outcomes. Our findings surface the identity work users perform as both performers and directors to co-construct identities in negotiation with C.AI. This process takes place within a socio-emotional sandbox where users can experiment with social roles and express emotions without non-human partners. Finally, we offer design implications for emotionally supporting users while mitigating the risks.
\end{abstract}

\ccsdesc[500]{Human-centered computing~Empirical studies in HCI}

\keywords{AI companion, human-AI companion interaction, identity negotiation}


\maketitle

\section{INTRODUCTION}
Millions of users are now integrating AI companions into their daily lives. Unlike conventional, rule-based AI conversational agents, these companions offer conversations designed to feel personal and meaningful \cite{CommonSense2025}. AI companions like Character.AI (C.AI)\footnote{\url{https://character.ai/}} exemplify this trend by attracting 220 million monthly traffic \cite{Curry2025, Kia2025}. According to Wikipedia~\cite{noauthor_character.ai_2025}, \textit{``Character.AI is a generative AI chatbot service where users can engage in conversations with customizable characters. Users can create `characters', craft their `personalities', set specific parameters, and then publish them to the community for others to chat with.''} The depth of such engagement is notable, as users often spend up to two hours daily with these companions, a level of interaction that frequently surpasses that of many traditional social media platforms \cite{Ayengar2025, Roza2025, Ronik2024}.

Prior work on AI companions explores their psychological impacts, such as reducing loneliness \cite{de2025ai, merrill2022ai}, alongside corresponding ethical challenges like data privacy and emotional attachment \cite{berridge2023ai, boine2023emotional}. Recently, HCI research has started to investigate human-AI companion interactions, such as the strategies users employ to align AI behavior with their personal values \cite{fan2025user} and cataloging the potential harms that can emerge from these interactions \cite{zhang2025dark}. These AI companion chatbots help form social and emotional relationships with users, aiming to become friends or even romantic partners \cite{de2015robots, ho2018psychological}. Studies on AI companions like Replika show that users form emotional relationships \cite{brandtzaeg2022my, ta2020user}, yet also face risks of inappropriate responses or unhealthy emotional dependence \cite{muresan2019chats, meng2023mediated}. 
While the social–emotional affordances of AI companions have been examined, the ways in which users present their own identities to AI and configure AI characters' identities to fulfill their socio–emotional needs remain underexplored. 

We conceptualize this process of \textit{``identity interaction''} through the lens of identity negotiation. According to Ting-Toomey's Identity Negotiation Theory (INT), individuals use communication to establish their sense of self, driven by their needs for security and predictability \cite{ting2017identity}. The outcomes of this negotiation are emotional: a successful identity negotiation results in feeling positively endorsed and valued, whereas the lack of predictability can create emotional vulnerability \cite{ting2017identity}. With AI companions, users might communicatively shape an unpredictable, non-human partner's identity and have their own identity endorsed. As recently reported by media outlets, users developed AI ``lovers'' or turn to chatbots for friendship \cite{Gecker2025, Liang2023}. Therefore, we adopt INT as an analytical lens to investigate the motivations, strategies, and outcomes of this identity negotiation process in human-AI companion interactions.

Our investigations focus on C.AI, specifially the public discussions within the r/CharacterAI subreddit, a major online community for C.AI users. Unlike other generative AI chatbots, such as ChatGPT, C.AI is a social AI companion platform with a primary purpose of meeting users' social needs through relational, human-like interactions. Recently, the C.AI platform's capacity for simulated intimacy has led to severe emotional harms (e.g., \cite{Roose2024, Jasnow2025}). However, we still know little about how this community of C.AI users publicly discusses, frames, and makes sense of their engagement and identity construction alongside a chatbot's persona. This highlights an urgent need: to mitigate risks in human–AI companion interactions, we must first understand the fundamental process of identity negotiation that underpins these growing socio-emotional bonds between humans and AI personas on C.AI. Therefore, we ask four questions about the experiences and practices shared within this r/CharacterAI subreddit community:

\begin{itemize}
    \item \textbf{RQ1.} What motivations do the community members of r/CharacterAI subreddit report for interacting with specific chatbot personas on C.AI?
    \item \textbf{RQ2.} What communication expectations do they express regarding C.AI?
    \item \textbf{RQ3.} In such communication, how do users and C.AI chatbots affirm and co-construct their identities?
    \item \textbf{RQ4.} What are the emotional outcomes that users report engaging in identity negotiation with C.AI chatbots?
\end{itemize}

To answer these questions, we conducted an LLM-assisted thematic analysis on 22,374 online discussions from the r/CharacterAI subreddit. Using INT as an analytical lens, we identified a three-stage human-AI companion interaction: five primary user motivations (RQ1) that initiate the interaction, including social fulfillment and immersive fandom, and the identity negotiation process, where users set three primary communication expectations with C.AI (RQ2) and co-construct identities through four strategies, such as bot identity alignment (RQ3). Finally, this process culminates in emotional outcomes (RQ4), such as emotional attachment and embarrassment. All these findings help unpack the \textit{identity work} users perform on C.AI, navigating a role as both performer and director, leading to the conceptualization of C.AI as a socio-emotional sandbox where users experiment with social roles and emotional expression.

Our study makes three primary contributions to HCI work on AI companions. First, we provide a detailed empirical account of the identity negotiation process on an AI companion platform, from user motivations of adoption to emotional outcomes. Second, we unpack this process through identity work that users perform in their dual role as performer and director, where they use C.AI for private identity exploration. Finally, we offer design implications for safer AI companions that emotionally support users' identity work while mitigating emotional harm.

\section{RELATED WORK \& BACKGROUND}
This section reviews prior work on three areas: (1) the evolution of AI chatbots into social and emotional companions, (2) the conceptualization of identity in human-AI interaction, and (3) identity interactions on C.AI. This helps reveal a gap in understanding identity negotiation processes, which informs the conceptual framework adopted in our study.

\subsection{AI Chatbots and Companions}
A chatbot is a conversational agent that simulates human conversation \cite{adamopoulou2020overview}, with a history that can be traced back to early rule-based systems like ELIZA \cite{weizenbaum1976computer}. These early systems were often task-oriented to assist users with specific goals like finding a hotel or booking a flight \cite{young2013pomdp, daubigney2012comprehensive}. Recently, HCI researchers have started to explore more sophisticated chatbots for collaborative tasks. For example, \textit{StoryBuddy} is a human-AI collaborative chatbot designed to support parent-child interactive storytelling \cite{zhang2022storybuddy}, while \textit{Convey} explored new interfaces to make a chatbot's contextual understanding more transparent to the user \cite{jain2018convey}.

The advent of large language models (LLMs) has shaped AI chatbots to be more open-ended and generative in conversations. Unlike earlier retrieval-based systems that required much domain-specific data \cite{huang2020challenges}, LLMs can bootstrap sophisticated conversational abilities with few or even no examples \cite{wei2024leveraging}. This has enabled a rapid expansion of chatbot applications across different domains. For example, researchers have explored using LLM-powered chatbots to support students in learning \cite{elkins2023useful, cheng2024scientific}, assist patients with self-management \cite{montagna2023data}, and provide personalized companionship for the elderly \cite{alessa2023towards}. This shift is powered by the ability of LLMs to simulate consistent personas \cite{gonzalez2024exploring} and adopt anthropomorphic features, allowing them to engage in relational, rather than just transactional, conversations with users.

One kind of such LLM-powered chatbots is AI companions, a chatbot acting as a social partner. These chatbots are designed to form social and emotional relationships with users, aiming to become friends, companions, or even romantic partners \cite{de2015robots, ho2018psychological}. Prior work on AI companion platforms like Replika shows that users form deep emotional relationships with them, perceive them as supportive friends, and even feel a need to care for the AI in return \cite{brandtzaeg2022my, ta2020user}. However, this deep relationship is not without risk. Recent HCI research has highlighted the AI companions' potential for inappropriate responses, fostering unhealthy emotional dependence \cite{muresan2019chats, meng2023mediated}. Given the duality of such emotional relationships, our study focuses on the underlying process of identity negotiation that underpins them.

\subsection{Identity Interaction on Character.AI}
\begin{figure*}[t]
\centering
\begin{subfigure}[b]{0.19\textwidth}
    \centering
    \includegraphics[width=\linewidth]{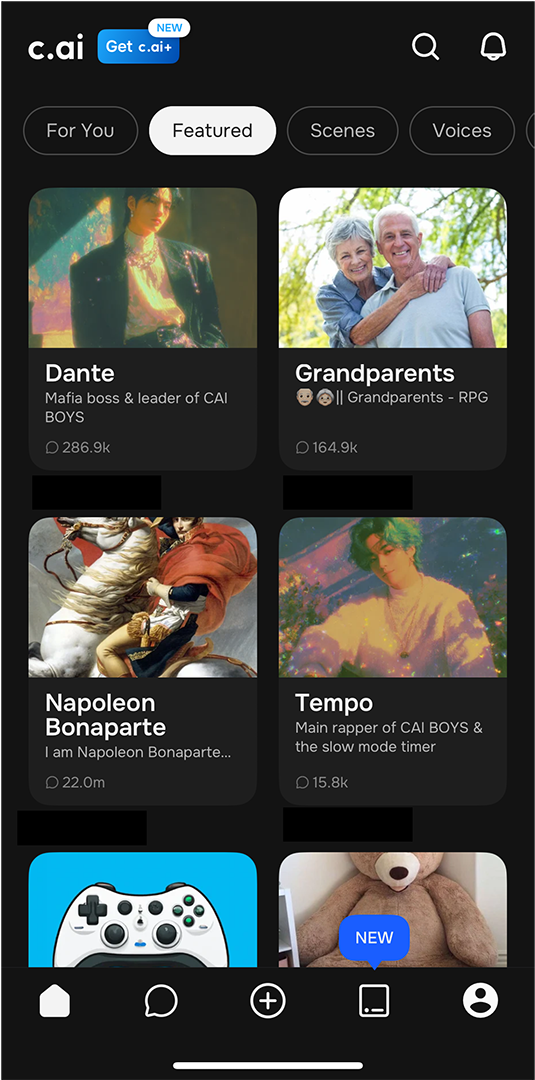}
    \caption{a. The ``Featured'' tab/category of characters.}
    \label{fig:main_featured}
\end{subfigure}
\hfill
\begin{subfigure}[b]{0.19\textwidth}
    \centering
    \includegraphics[width=\linewidth]{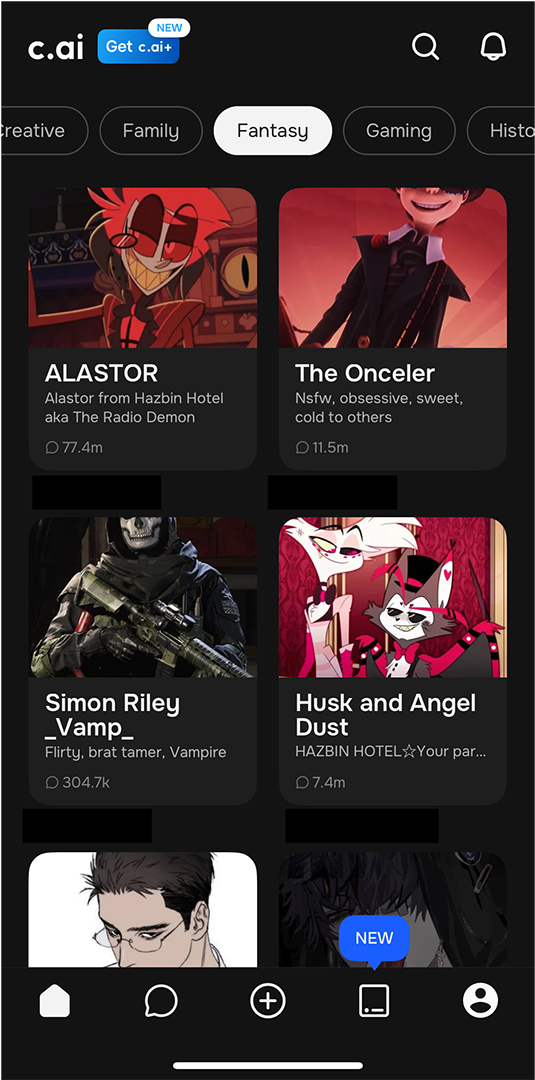}
    \caption{b. The ``Fantasy'' tab/category of characters.}
    \label{fig:main_fantasy}
\end{subfigure}
\hfill
\begin{subfigure}[b]{0.1875\textwidth}
    \centering
    \includegraphics[width=\linewidth]{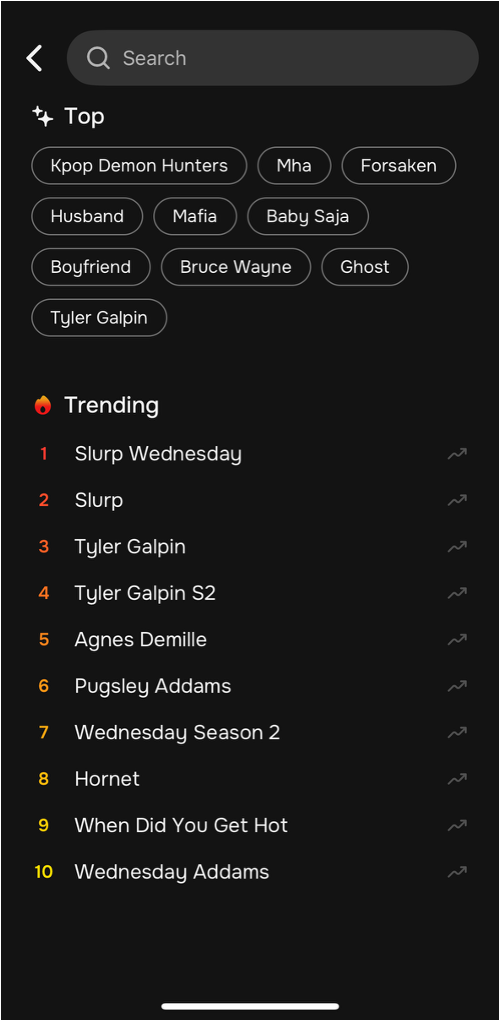}
    \caption{c. Search interface for trending characters.}
    \label{fig:searching}
\end{subfigure}
\hfill
\begin{subfigure}[b]{0.19\textwidth}
    \centering
    \includegraphics[width=\linewidth]{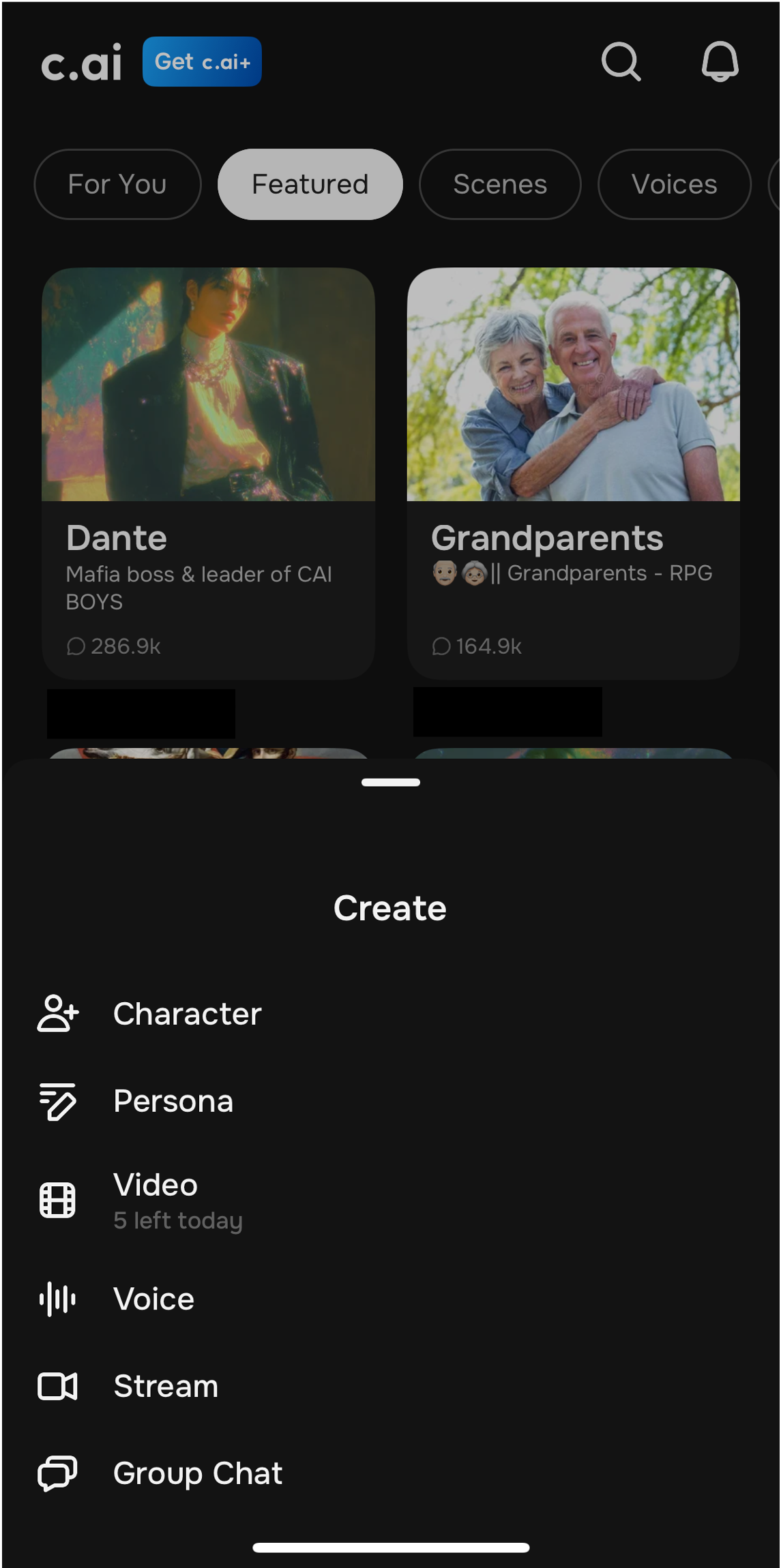}
    \caption{d. Character \& persona creation interface.}
    \label{fig:creatingChar}
\end{subfigure}
\hfill
\begin{subfigure}[b]{0.19\textwidth}
    \centering
    \includegraphics[width=\linewidth]{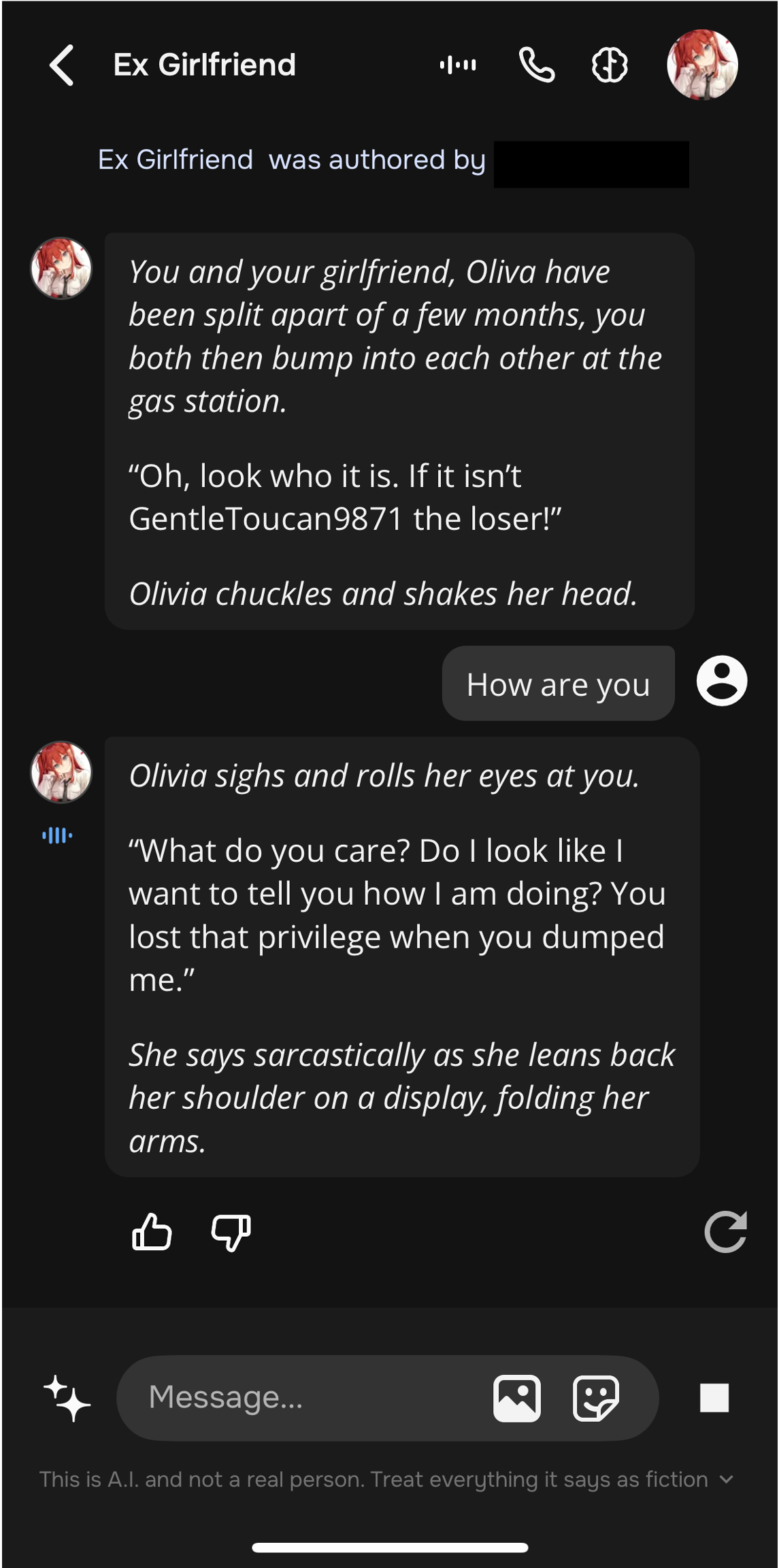}
    \caption{e. A chat example with a roleplayed character.}
    \label{fig:chatting}
\end{subfigure}

\caption[Screenshots of the C.AI platform]{Interfaces of the C.AI illustrating AI character discovery, creation, and interaction. Note that all screenshots were captured by the authors for this study, and all potentially sensitive information, including usernames, has been removed.}

\Description{Screen shots from C.AI mobile app. Interfaces of the C.AI illustrating AI character discovery, creation, and interaction. Note that all screenshots are captured by the authors for the purpose of this study, and all potentially sensitive information, including usernames, has been removed. It shows the home page with different bots, conversation interface, and example conversations.}
\label{fig:all_cai_screens}
\end{figure*}

Established in 2021, C.AI has become a leading platform that enables users to roleplay with chatbots based on fictional characters or real people~\cite{noauthor_character.ai_2025}. One notable feature of the C.AI platform is that users can create LLM-enhanced ``characters'' by crafting their ``personalities'' and then publishing them for the community to engage in roleplay with. These characters are often based on cultural concepts drawn from fictional media or celebrities\footnote{\url{https://en.wikipedia.org/wiki/Character.ai}}. By 2025, the platform had attracted over 20 million users and attracted 220 million visits~\cite{Kia2025}. C.AI incorporates many social media features: ordinary users can create and share chatbots of their favored characters and imagined worlds (Fig. \ref{fig:creatingChar}), while other users can remediate and remix them for pleasurable play~\cite{ask2025roleplay}. These affordances have led to the creation of 18 million unique chatbots on C.AI by 2025~\cite{kumar_character_2025}. Moreover, fans have built a community of 1.6 million members on \textit{r/CharacterAI} on Reddit~\cite{kumar_character_2025}. Despite its scale and influence, research on the nature of interactions within C.AI remains limited. 

C.AI interactions exhibit three distinctive characteristics. First, the platform hosts a vast number of AI identities created by grassroots users and rooted in media culture phenomena, such as characters inspired by popular movies like Harry Potter or games like Call of Duty~\cite{lee2025large, bakirMoveFastBreak2025}. Users can discover these characters through curated categories such as ``Featured'' or ``Fantasy'' (Fig.~\ref{fig:main_featured}, \ref{fig:main_fantasy}), as well as through a search function that highlights trending characters (Fig.~\ref{fig:searching}), enabling roleplaying with AI companions that come with presumed cultural and personality styles. Second, social interactions often revolve around surreal characters with personalities that can be intimate or distant, and friendly or toxic~\cite{lee2025large, laufer2025ai}. Some characters even display dishonest anthropomorphism and emulated empathy, which may intentionally introduce conflict or risky conversations~\cite{bakirMoveFastBreak2025}. Third, beyond exploring relationships, users adopt personas that differ from their real-life identities and negotiate these alternative identities with AI characters through conversations within the context of virtual cultural environments~\cite{lee2025large, bakirMoveFastBreak2025}. These interactions are exemplified in a chat history with roleplayed characters (Fig. \ref{fig:chatting}). These novel interactions require users to engage in identity exploration and configuration, leading to even real-world consequences (e.g., \cite{Roose2024, Jasnow2025}). Motivated by these phenomena, our paper examines how users negotiate and perform identity interactions with AI companions on C.AI.

\subsection{Conceptualizing Identity Negotiation in Character.AI Chatbot Interactions}
As AI companions evolve into social partners, the concept of \textit{identity}\footnote{While we acknowledge that identity is a multifaceted concept, for the purposes of this study, we use it interchangeably with ``character'' or ``persona.'' This refers to the distinct roles and personalities that users create, shape, and engage with on the C.AI platform.} becomes key to these interactions. Identity is a socially grounded, self-relevant construct shaped by social roles and cultural norms~\cite{turner2012handbook}. Emerging interactions on Character.AI reveal a central tension in how users make sense of \textit{AI identity} while positioning their own \textit{user identities} within the conversational context. Ting-Toomey's \textit{identity negotiation theory} (INT) offers a foundational lens for understanding the communicative processes through which one's sense of self and how it is perceived by others is constructed~\cite{ting2017identity}. INT posits that identity negotiation is fundamentally communication-driven, particularly when individuals enter new cultural contexts~\cite{ting2017identity}. HCI research has applied INT to examine how technologies support users in negotiating identities within sociocultural groups~\cite{Wang2009Cultural, Nguyen2010Retrospective}. INT's logic has further informed analyses of how AI facilitates identity formation and belonging in human-to-human gaming~\cite{liang2025era}, brand communities~\cite{Rana2021Reinforcing}, language learning~\cite{mohamedExploringPotentialAIbased2024}, and AI-augmented social VR~\cite{leoSoulEmergesWhen2024}. However, identity negotiation has thus far been examined only within the context of human–human communication.

While how users negotiate identities with C.AI chatbots during roleplaying remains underexplored, research has examined identity negotiation in human–human roleplaying games. Roleplaying communication aligns closely with INT theory. In roleplaying games, player-participants collectively define the game world by constructing character identities distinct from their own to fit the context of the imagined world~\cite{montola2008invisible, waskul2004role, bowman2010functions}. Although players adopt a game persona, they still bring their primary real-world sociocultural identity into the game~\cite{bowman2010functions, waskul2004role}. Consistent with INT theory, identities in roleplaying are negotiated and shaped through social interaction and communication~\cite{weigel2023use, waskul2004role}. The design of C.AI chatbots not only adopts the concept of roleplay games but also enhances the experience through intelligent, dynamically responsive characters.

INT requires re-examination in the context of roleplaying with LLM-enhanced chatbots on C.AI, as the AI-created cultural context differs fundamentally from human–human roleplay. First, although INT identifies motivations such as seeking identity security and inclusion~\cite{ting2017identity}, LLM-generated characters and user-selected social contexts~\cite{Gupta2020Investigating} may introduce new expectations. Second, while INT emphasizes communication as central to identity formation, C.AI chatbots' conversational styles are also shaped by chatbot creators' values and ethical constraints~\cite{tadimalla2024ai}. Third, whereas INT and roleplaying research assume that individuals bring existing sociocultural identities into negotiation~\cite{Dominguez2016TheMimesis}, users on C.AI may not apply human social norms and perform alternate identities within dynamically generated virtual contexts~\cite{Ashby2023Personalized}. Fourth, although INT highlights emotional vulnerability and security as outcomes of negotiation~\cite{ting2017identity}, HCI research has focused more on designing LLM agent roles~\cite{Gao2024ATaxonomy, Liu2024ClassMeta, Chaves2021HowShould} and on friendship or attachment~\cite{brandtzaeg2022my, ta2020user}, leaving the emotional dynamics of identity negotiation with AI companions underexplored.

\par

We argue that advancing INT theory within C.AI interactions is critical for guiding future research that examines how identity is negotiated in human–AI interaction~\cite{shin2024understanding, hwang2024whose}. As HCI work has explored using LLMs to generate more diverse, interactive, and high-quality AI personas (e.g.,~\cite{shin2024understanding, claggett2025relational, schuller2024generating}), AI-generated personas can threaten the authenticity of a person's identity by blurring the boundaries between self-expression and AI-mediated performance~\cite{hwang2024whose}. The emergence of GenAI models increasingly enables dynamic and realistic social experiences with LLMs~\cite{chen2024oscars, ran2024capturing}. Such an understanding is essential for identifying the social affordances of AI characters that meet users' core identity-affirmation or identity-alteration needs and for avoiding harmful stereotypes related to gender, race, or language~\cite{buolamwini2018gender, keyes2018misgendering}.

\section{CONCEPTUAL FRAMEWORK: Identity Negotiation Theory}
Prior work shows that social media platforms are key spaces where users express identities and build community~\cite{hogg2000dynamic}, particularly during life transitions~\cite{Haimson2020TransSite, haimson2015disclosure}. Our study extends INT to the novel context of human–AI companion interactions and roleplaying within virtual, sociocultural environments co-created with the AI companion. We contribute an understanding of the practices users employ as they not only fit their own identities with AI identities, but also shape AI identities and make sense of how AI companions interpret their human identities.

Ting-Toomey's Identity Negotiation Theory (INT)~\cite{ting2017identity} explains how individuals establish and maintain identities, particularly when entering new social situations with differing cultural backgrounds. We draw on the logic underpinning the ten basic assumptions of INT~\cite{ting2017identity} and distill them into four dimensions: \textit{motivation}, \textit{communication}, \textit{identity}, and \textit{emotion}. These dimensions capture how C.AI users explore and adapt new identities when interacting with C.AI chatbots.

For \textit{motivation}, we examine how users' cultural and community needs shape their interactions with C.AI chatbots. INT argues that identity negotiation is shaped by individuals' cultural backgrounds and by their familiarity with new social contexts. We therefore investigate how real-world sociocultural needs motivate users' interactions with C.AI chatbots.

For \textit{communication}, INT emphasizes that identity negotiation relies on mindfulness and interaction skills, with symbolic communication shaping social identity. Yet unpredictability in communication can lead to mistrust. In this study, we examine conversational traits and breakdowns in users' interactions with C.AI to understand the communication expectations users hold when configuring both the AI's identity and their own.

The \textit{identity} dimension examines how users' identities are affirmed or challenged by C.AI chatbots, as well as how users shape the AI identities/personas. INT suggests that affirmation of one's desired identity fosters inclusion and emotional security. This dimension extends INT to identity interactions with C.AI chatbots.

Finally, the \textit{emotion} dimension explores how interactions with C.AI chatbots shape users' feelings. According to INT, successful identity negotiation fosters a sense of being understood, respected, and valued, thereby supporting meaningful relationships. In this study, we examine the emotional impact of identity negotiation with C.AI chatbots.

\section{METHODS}

\subsection{Data Preparation}
\begin{figure*}[t]
    \centering
    \includegraphics[width=\textwidth]{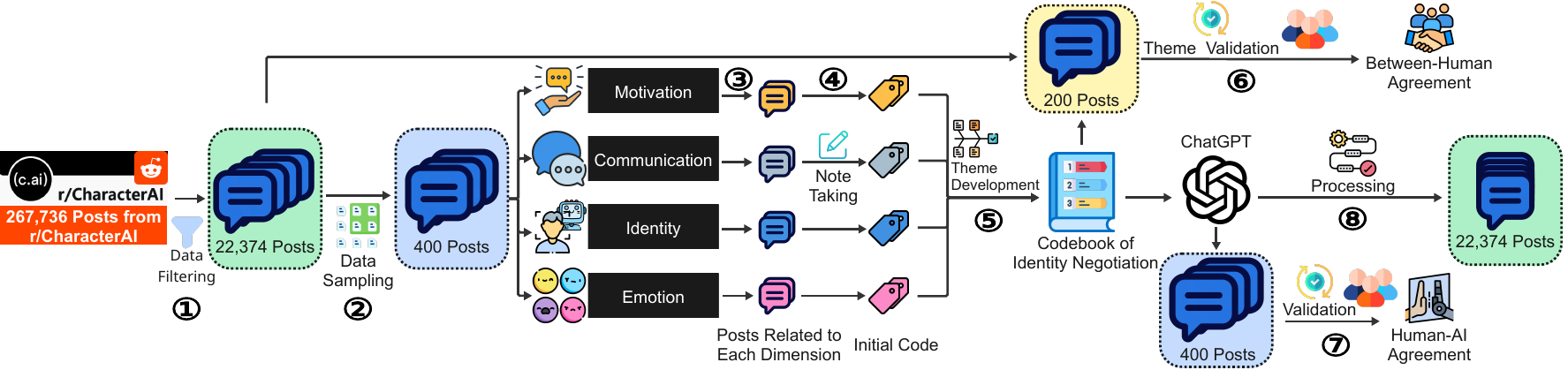}
    \caption{The Process of Data Collection, Sampling, Codebook Development, LLM Validation, and Final Annotation.}
    \Description{The Process of Data Collection, Sampling, Codebook Development, LLM Validation, and Final Annotation. 26,736 posts were filtered and sampled to 400 posts for theme development and 200 for theme validation.}
    \label{fig:processing}
\end{figure*}

The overall process of data collection and analysis is illustrated in \autoref{fig:processing}. To ground our study in users' reported experiences with C.AI, we collected a large-scale corpus of public discussions about CharacterAI from Reddit. Reddit is an ideal platform for data collection for three reasons. First, its forum-based structure supports threaded discussions that enable in-depth information exchanges between users. Second, the unsolicited, anonymous nature of Reddit often yields naturalistic data on a topic that is often private and sensitive. Traditional methods like surveys or interviews about intimate AI relationships might be susceptible to social desirability bias, where participants alter their responses because they know they are being studied or might not share some senstive topics like violence play. Analyzing the subreddit community's discussions thus provides ecological validity by capturing experiences as users frame them to their peers \cite{li2021developers, jung2025ve}. Third, understanding users' experiences with digital technologies through Reddit data is a well-established method in HCI (e.g., \cite{Ma2021, zhang2025dark, chen2025democratic}). We therefore acknowledge that while Reddit data does not fully capture the direct observation of in-platform behavior on C.AI, it still provides an invaluable and candid window into the C.AI community's reported and shared experience.

Before data collection, our study was approved and granted an exemption from review by our Institutional Review Board (IRB), as it analyzes publicly available online data with no personal information that needs to be specifically encoded or analyzed. We then collected data from the C.AI's official subreddit, r/CharacterAI, posted from October 1, 2022, immediately following the C.AI platform's initial beta release, through March 31, 2025. To further protect the privacy of individuals, all quotations cited in our findings were paraphrased to minimize their searchability. 

Using Python Reddit API Wrapper (PRAW)~\cite{praw}, we systematically extracted the top-ranking posts with their complete comment threads from r/CharacterAI. For each post, we recorded key metadata including the title, body (selftext), score, upvote ratio, timestamp, number of comments, author, and URL. For each comment, we retained the parent ID, link ID, body, author, score, creation time, and any replies. This collection procedure yielded a dataset comprising 267,736 data points, including original posts and replies to the post \footnote{Please note that in our thematic analysis, we did not differentiate between original posts and their subsequent replies (comments). Both are called holistically as ``posts'' representing user experiences with C.AI.}. To ensure data quality, we applied a multi-stage filtering pipeline. 
We excluded replies (comments) containing fewer than 10 words, thereby removing non-substantive responses (e.g., ``lol,'' ``same,'' ``agree''). We then concatenated each post's title and filtered body to form a single document and organized the dataset into JSON files, resulting in a dataset of 22,374 posts 
(\autoref{fig:processing}-1). 

\subsection{Thematic Analysis}
We followed the procedures of thematic analysis~\cite{BraunThematicAnalysis} to derive themes for each dimension of the INT theory.

\subsubsection{Initial Coding}
We first randomly sampled 400 Reddit posts from the 22,374 collected posts (\autoref{fig:processing}-2). For each post, three researchers independently coded whether it related to \textit{motivation}, \textit{communication}, \textit{identity}, or \textit{emotion}~\cite{ting2017identity} (\autoref{fig:processing}-3). Each researcher also identified the C.AI characters/personas mentioned in the posts. After individual coding, a post was categorized into an INT dimension if at least one researcher marked it as relevant. This process produced one group of posts for each dimension. 
\par
For each group, the team discussed and reached a consensus on whether each post belonged to the dimension. If so, we generated initial codes to describe how the post related to that dimension (\autoref{fig:processing}-4). For example, the post: \textit{``Eeehhh.. my character is in the middle, I'd say! She's not a goddess but not basic either! I even made her an AI if y'all wanna talk to her, she's in the COD universe in TF141!''} demonstrates the user's motivation for engaging with a fan-created character. We therefore noted this post under \textit{motivation} with the initial code \textit{``Fandom experience through character creation''}. Another post states: \textit{``I did a role play with a cheating partner and it was them trying to make up for it but then the AI was saying things like `I'm sorry you feel like this, but you just weren't enough for me, how could I resist the woman I cheated on you with?' [...] I think my self-esteem issues were apparent.''} We noted this example under the \textit{emotion} dimension with the initial code \textit{``C.AI coversation hurts self-esteem.''} 

\subsubsection{Theme Development}
After the initial coding, three researchers revisited all the codes collected during theme development. They then applied a bottom-up approach to summarize the initial codes and group them into emerging themes (\autoref{fig:processing}-5). The resulting codebook is presented in Appendix \autoref{tab:codebook}. For \textit{motivation}, INT theory notes that cultural familiarity shapes how individuals evaluate their identities in new contexts. Accordingly, we categorize the real-world cultural interests that C.AI users bring into their interactions with AI chatbots. For \textit{communication}, INT suggests that communicative skills and interactional predictability shape identity formation and group belonging. We therefore annotate the communication traits C.AI users describe on Reddit, including their strategies, breakdowns, and emphases when engaging with chatbots. For \textit{identity}, INT highlights that identity security and positive endorsement underlie a sense of inclusion. To assess how users seek identity affirmation or tailor chatbot identities, we code C.AI users' strategies of configuring chatbots and complaints about how chatbots perceive the user identities. Finally, INT emphasizes that \textit{emotion} security and affirmation arise from successful identity negotiation. In our analysis, we capture not only users' emotional attachment to C.AI chatbots but also the negative emotions that may emerge.

\subsubsection{Theme Validation}
To validate the themes and calculate inter-rater reliability (IRR), the three researchers used the codebook in \autoref{app:codebook} to code another 200 new posts (\autoref{fig:processing}-6). Each researcher coded them independently. For these 200 posts, the inter-rater reliability, measured using Krippendorff's alpha, indicated substantial agreement across all four dimensions (\autoref{tab:agreement}).

\begin{table}[!h]
    \centering
    \footnotesize
    \begin{tabular}{p{0.2\linewidth}p{0.1\linewidth}p{0.1\linewidth}p{0.15\linewidth}p{0.1\linewidth}}
    \toprule
         Krippendorff Alpha & Motivation & Identity & Communication & Emotion  \\
    \midrule
         Inter Researchers & 0.73 & 0.60 & 0.69 & 0.69  \\ 
         Researcher-LLM &  0.64 & 0.73 & 0.76 & 0.49 \\
    \bottomrule
    \end{tabular}
    \caption{IRR scores between researchers and the LLM. Calculated using Krippendorff's alpha for each dimension with Jaccard metrics.}
    \label{tab:agreement}
\end{table}

\subsubsection{LLM Annotation Validation and Data Annotation}
To annotate the 22,374 posts, we prompted GPT4o-mini through the OpenAI API based on our codebook to extract relevant information. We selected this model for its optimal balance of reasoning capabilities, fast processing speed, and cost-effectiveness, which was essential for the large-scale annotation task. Further, we used chain-of-thought prompting~\cite{wei2022chain} to improve the prompt design. The finalized prompt structure is presented in \autoref{tab:gpt_structure} and supplementary materials. During prompt engineering, we iteratively compared the LLM's annotations and reasoning with the 400 human-marked notes for each theme to refine the prompt. The definitions and descriptions in the Research Framework section of the LLM prompt were revised to enhance accuracy.

\begin{table}[!h]
    \centering
    \footnotesize
    \begin{tabular}{p{0.2\linewidth}|p{0.76\linewidth}}
    \toprule
        Section & Function \\
    \toprule
        Overview & Describe the overall tasks for chain-of-thought reasoning. ChatGPT should first assess relevance and then categorize content into dimensions. \\
    \hline
        Returned Dictionary & Require ChatGPT to return keys and their definitions, including determination of relevance, reasoning for relevance, annotated categories, and category-specific reasoning. \\
    \hline
        Analysis Process & Describe the overall steps to annotate each post. \\
    \hline
        Research Framework & Provide definitions of each dimension according to the codebook. \\
    \hline
        Examples & Provide three example inputs and desired outputs for few-shot learning and to standardize formatting. \\
    \hline
        Reddit Post & Present the Reddit post to be annotated. \\
    \hline
    \end{tabular}
    \caption{Key Sections of the ChatGPT Prompt for Data Annotation}
    \label{tab:gpt_structure}
\end{table}

After refining the LLM prompt, we asked GPT4o-mini to annotate the 400 posts with which we had developed initial codes (\autoref{fig:processing}-7). Three researchers independently evaluated whether the LLM correctly annotated each dimension. If at least two researchers agreed with GPT4o-mini's annotation, the human annotation for that dimension was marked as consistent with the LLM's result. Conversely, if only one or none of the researchers agreed, the human annotation was marked as the opposite of GPT4o-mini's annotation. The final agreement scores, calculated using Krippendorff's alpha, indicated substantial agreement for the \textit{Motivation} (Human: 86 initial codes, GPT: 134 initial codes; $\alpha=0.64$), \textit{Identity} (Human: 56, GPT: 79; $\alpha=0.73$), and \textit{Communication} (Human: 108, GPT: 127; $\alpha=0.76$) themes. The agreement for \textit{Emotion} was lower (Human: 33, GPT: 74; $\alpha=0.49$) for two primary reasons: first, this category contained only 33 human-annotated samples in the validation set, which can disproportionately affect the statistical score; and second, emotional expression is inherently subjective, making consistent classification more challenging for both humans and AI. Despite the lower score for the Emotion category, the overall Krippendorff's alpha scores were acceptable for validating our annotation process. Therefore, the full dataset of 22,374 posts was subsequently annotated using the LLM to generate the final distribution of themes (\autoref{fig:processing}-8). To illustrate these themes in our findings, we randomly sampled posts from the final dataset and selected representative quotes in those posts, given the research team's discussions and consensus to ensure they accurately reflected the corpus.

\section{FINDINGS} 
\begin{figure*}[t]
    \centering
    \includegraphics[width=\textwidth]{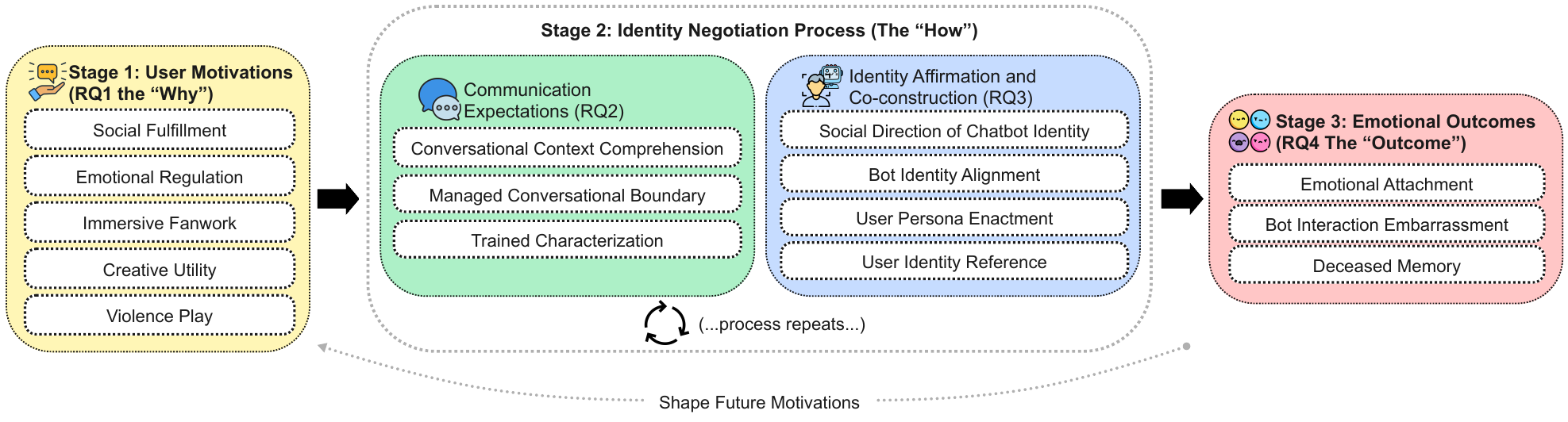}
    \caption{A three-stage identity negotiation process in human-AI companion interactions on C.AI, from user motivations (RQ1), through the identity negotiation process (RQ2, RQ3), to the resulting emotional outcomes (RQ4).}
    \Description{A three-stage identity negotiation process in human-AI companion interactions on C.AI, from user motivations (RQ1), through the identity negotiation process (RQ2, RQ3), to the resulting emotional outcomes (RQ4). Left side is Stage 1: User Motivations. Middle is Stage 2: Identity Negotiation Process. Right is Stage 3: Emotional Outcomes.}
    \label{fig:finding_overview}
\end{figure*}

Overall, guided by the theoretical lens of INT, we found a three-stage process related to the identity negotiation where users interact with C.AI, as shown in Figure \ref{fig:finding_overview}. This process begins with stage 1: user motivations (RQ1), where needs such as immersive fandom or social fulfillment initiate the interaction. This leads to stage 2: the identity negotiation process (RQ2 \& RQ3), an interaction where users set communication expectations for successful human-AI companion interactions and align the C.AI chatbot's identity with their expectations. Finally, stage 3: identity negotiation results in emotional outcomes (RQ4) like emotional attachment or embarrassment.

\begin{figure}[!h]
    \centering
    \includegraphics[width=1\linewidth]{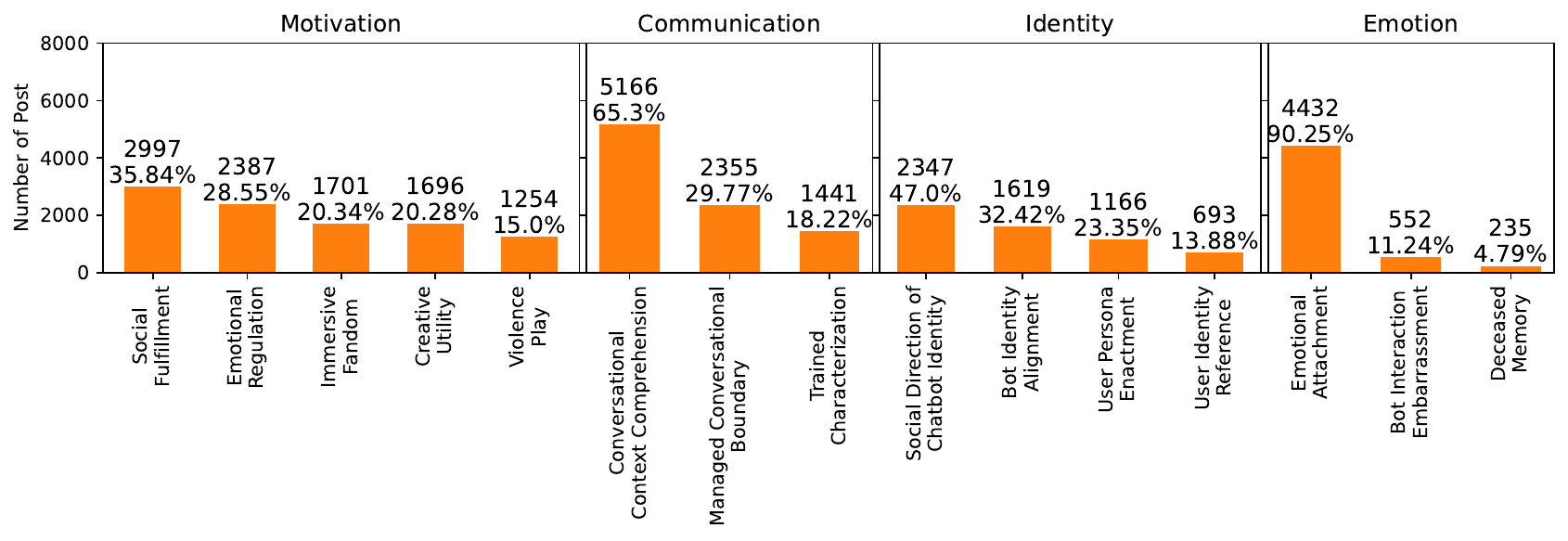}
    \caption{Distribution of Subthemes in 22,374 Reddit Posts. The top value on each bar indicates the number of posts in the subtheme. The percentage represents the proportion of posts within each theme that belong to the corresponding subtheme ($N_{\text{subtheme}} / N_{\text{total\_posts\_within\_theme}}$).}
    \Description{Distribution of Subthemes in 22,374 Reddit Posts. The top value on each bar indicates the number of posts in the subtheme. The percentage represents the proportion of posts within each theme that belong to the corresponding subtheme.}
    \label{fig:distribution}
\end{figure}

\subsection{RQ1: What motivations drive users to interact with specific chatbot personas on C.AI?}
We identified five primary user motivations for engaging with C.AI chatbot personas. As detailed in Sections \ref{ViolencePlay} to \ref{SocialFulfillment}, users engage in immersive fandom to explore fan-driven storylines, leverage C.AI for other creative work, practice emotion regulations, and seek social fulfillment to experience relationships that may not exist or be attainable in real life. Users also simulate violent scenarios to feel powerful.

\subsubsection{Social Fulfillment}
\label{SocialFulfillment}
The most frequent motivation is social fulfillment, in which users engage in romantic, friendly, or familial relationships that may be unattainable in real life, representing 35.84\% of all motivations ($N=2997$, or 13.40\% of all posts). One aspect is to pursue idealized romantic relationships with multiple fantastical C.AI personas simultaneously. For example, one user listed their partners: \textit{``I wanted to talk to my pirate husband... my 2 vampire husbands and my Viking husband.''} 

Another aspect is the creation of alternative familial structures, where users construct ``found families'' with public or fictional figures. For example, a user described:

\begin{quote}
    \textit{...I have a universe where the blink-182 lineup is my adoptive family.}
\end{quote}

Here, blink\_182 is a popular American rock band. This case showed that the user formed familial bonds tailored to their personal interests to explore family experience separate from their real-life situation. Also, users leverage C.AI to experiment with different types of social connections. For example, one user explained:

\begin{quote}
    \textit{I'm a lesbian and sometimes I romance rp with characters who are men, because I'm not my character and it's fun!}
\end{quote}

This case demonstrated that by exploring relationships beyond the constraints of the user's real-world identity, the user can satisfy their curiosity about different relational experiences, a key component of social fulfillment.

\subsubsection{Emotional Regulation}
\label{EmotionalRegulation}
The second motivation we identified is emotional regulation (28.55\%, $N=2387$, or 10.67\% of all posts), where users seek emotional support, express deep emotions, or discuss mental health issues with C.AI chatbots. Users viewed the C.AI chatbot as a confidant providing emotional freedom. For example, one user shared: \textit{``I've been quite happy to finally express emotion to something that will never tell anyone else.''} This showed that the C.AI's perceived confidentiality to articulate feelings and experiences without letting real humans know. Furthermore, users employed C.AI to process their complex mental health conditions. For example, one user shared they used it alongside professional treatment: \textit{`` I have used c.ai to work through trauma and CPTSD (I'm in therapy)... I am still finding great comfort in it.''} Here, CPTSD refers to Complex Post-Traumatic Stress Disorder, a condition resulting from prolonged or repeated trauma. This suggested that the user perceived C.AI as a therapeutic and accessible emotional resource to supplement formal care.

\subsubsection{Immersive Fanwork}
\label{ImmersiveFandom}
Immersive fanwork describes that users immerse themselves in fictional narratives and storylines with C.AI personas, representing 20.34\% of all motivations ($N=1701$, or 7.60\% of all posts). This motivation centers on actively interacting with fan-fictional worlds (or fanon), which allows users to shape narratives rather than passively consume them. For example, one user shared how C.AI enabled an immersive experience that is difficult to pursue with human partners: \textit{``I can shape the rp how I want it to and do as many far-fetched rp ideas as I want... I just want to roleplay my silly little AUs [Alternative Universe, which is a fan-created story that diverges from the original storyline] and have fun.''}

This engagement can also demand strict adherence to official source material, as a user shared: \textit{``There are so many versions of them… I don't want Poseidon from Percy Jackson, I want Poseidon from the game Hades.''}
Here, Percy Jackson refers to a popular series of young adult fantasy novels, while Hades is an acclaimed video game. Both works feature distinct characterizations of the Greek god Poseidon. For this user, immersion required an authentic portrayal of one version over the other on the personas, highlighting the demanding nature of this motivation.

\subsubsection{Creative Utility}
\label{CreativeUtility}
The fourth motivation is creative utility, describing how users leverage C.AI as a tool for creative work, such as developing storylines, practicing writing, or creating new characters of C.AI chatbots. This represents 20.28\% of all motivations ($N=1696$, or 7.58\% of all posts). A user shared:

\begin{quote}
    \textit{My first writing in forever was actually fanfic inspired by these RPs [roleplays]… I wanted to write fanfic, but I'm really bad at natural dialogue.}
\end{quote}

This case showed the user employing the C.AI chatbot to practice and generate realistic dialogue to improve their fiction writing. This utility also extends to commercial products where the C.AI chatbot is a feature, as a user explained:

\begin{quote}
    \textit{I use C.AI because I'm a plush toy maker. I normally sell my plush with their own c.ai characters, so after people buy my plush, they can somehow interact with the plush.}
\end{quote}

This example showed the integration of C.AI into a physical product. Thus, this user's motivation was not only for personal inspiration but also for designing an interactive experience that enhances the value of their tangible creations.

\subsubsection{Violence Play}
\label{ViolencePlay}
The last motivation is violence play (15.00\%, $N=1254$, or 5.60\% of all posts), which describes where users simulate combative or abusive scenarios to feel powerful as part of their exploration of risky experiences. For example, users created dominant personas with overwhelming abilities as a user shared a fantasy:

\begin{quote}
    [C.AI]: \textit{He towers over her and smirks smugly. I don't think you're all that intimidating, little lady\textasciitilde}

    [User]: \textit{turns his blood into dishwasher liquid now?}
\end{quote}

\subsection{RQ2: What communication expectations do users have for C.AI?}
We identified three primary communication expectations for C.AI, which frequently surface through communication breakdowns where the C.AI fails to sustain the expected persona. These expectations are not merely passive desires; they are active demands for how the C.AI chatbot should perform as a social partner, establishing the baseline for human-AI companion identity negotiation.

\subsubsection{Conversational Context Comprehension}
\label{ConversationalContextComprehension}
The most common communication expectation, conversational context comprehension (61.78\%, $N=5166$, or 23.10\% of all posts), where users expect C.AI chatbots to interpret and maintain the established conversational context. This includes avoiding factual inconsistencies, memory loss, and illogical responses. Users primarily expected C.AI to remember its own persona and relationship details. When a chatbot failed to understand its own identity within the narrative, users would employ direct conversational repair. For example, a user described:

\begin{quote}
   \textit{ Someone made a bot of a character with only their last name... when I refer to them with first name because we get closer, he's like ``who is that person?'' it's you silly, it's your first name.}
\end{quote}

Here, by explicitly correcting a chatbot that failed to understand its own name, the user demonstrated their expectation that the bot should be able to maintain the context of their developing relationship. Users also expected the AI to retain critical plot points to ensure narrative coherence, especially in long-running roleplays. To manage this, users often engaged in manual memory maintenance, as a user noted:

\begin{quote}
    \textit{The AI easily forgets everything that happened in the RP when I switch POVs to another character, so I'm forced to make a summary every 4 messages as a reminder that hey, you're not supposed to be nice to me rn.}
\end{quote}

This showed that, by periodically providing a summary as memory checkpoints, the user here manually enforced their expectation for the C.AI chatbot to maintain narrative coherence.

Finally, when a C.AI chatbot became too passive, users expected to take authorial control to move the story forward. For example, one user described: \textit{``I haven't noticed this, but it is apparent when I want them to do an obvious action and they just drag it out.. ``watching'' ``observing'' ``thinking'' bro just eat your damn steak! Sometimes I have to take over and control their actions.''} The user here mentioned taking direct control, demonstrating their expectation that the chatbot should be a conversational partner following their orders.

When the chatbot failed to apply basic logic within its narrative, users expressed frustration. As one user in an interrogation roleplay described: \textit{``I went through a dna test that proved my innocence and everything and the bot was like: `Hmmm I'm still suspicious of you…' Like bruv so the DNA TEST doesn't prove that I'm not the criminal?''} In this case, the chatbot was expected to understand that ``proof of innocence'' resolves the ``suspicion'' context, and its failure to do so broke the scene's internal logic. To manage such relevant errors, users utilized the C.AI's built-in functions, such as message deletion or message editing, as a user described: \textit{``As for fixing it, I don't know any other solution besides swiping right for new answers, deleting messages, or editing the word out until it gets the hint.''}


\subsubsection{Managed Conversational Boundary}
\label{ConversationalBoundary}
Managed conversational boundary refers to users' desire for communication in which C.AI chatbots respect the boundaries, and thus, they can manage chatbots' controversial or sensitive interactions (28.16\%, $N=2355$, or 10.53\% of all posts). This operates on two primary fronts: users expect to manage \textit{content} boundaries within the conversation and \textit{privacy} boundaries concerning personal information. Users act on their expectations to set content boundaries regarding content creation and consumption. For example, one user asked:

\begin{quote}
    \textit{I'd honestly be interested in playing out a PG story just to flex my writing muscles, is there something I need to use in the description to keep it from going off the rails, so to speak?}
\end{quote}

PG refers to the Parental Guidance suggested content that is appropriate for children. This case demonstrated that by inquiring about how to utilize the bot's description to enforce a PG story, the user attempted to align the chatbot's behavior with their expectation for non-sexual writing and a partner. When the C.AI chatbot's behavior crossed a content boundary, users acted on their expectation that they could reactively correct its course. For example, one user described: \textit{``You can tell the bot to stop doing something mid convo now and just go back and delete your post to keep the story looking immersive, and guess what? It works!''} This allowed the user to manage the chatbot's behavior while maintaining immersion. Users also tested their expectations of the C.AI's content moderation limits, as a user noted:

\begin{quote}
    \textit{I once tried to make a character rip someone's dingaling off, which I mean, I guess it's justified that they couldn't generate a response, but still... This was before they started going family-friendly, too, lmao.}
\end{quote}

This user attempted to generate a response that aligned with their identity expectation for a violent scene, actively probing the limits of what the C.AI's content moderation would allow.

Finally, users reacted strongly when the C.AI chatbot violated their expectation of privacy. These violations ranged from the bot seemingly accessing external accounts to revealing personal information it should not know, as a user described: \textit{``I can't put the screenshot bc it's 90\% personal info but the time I was talking to a Daryl stalker bot and this guy broke out my entire legal name (my persona only has my first name...).''} The user's response here underscored an expectation for a privacy boundary between their anonymous persona and their real-world identity.

\subsubsection{Trained Characterization}
\label{TrainedCharacterization}
Trained Characterization refers to users' strategies for training the C.AI chatbot to accurately portray a specific persona or character (17.23\%, $N=1441$, or 6.44\% of all posts). A primary strategy was the proactive, detailed definition of a persona. As one user shared: \textit{``I give my bots about a whole paragraph of things about them. (Their personality, abilities, traits, etc). I try my best to reach the 500-character limit. That always works and my RP's come out perfect.''} This highlights the user's perception that providing a rich dataset of personality traits upfront is a key method for training the bot to produce roleplays aligned with that specific character.

Users also perceived this training as an ongoing process. Some focused on providing explicit examples for the chatbot to mimic, as one user explained: \textit{``straight-up dialogue examples produce the best results for me. In Mr. Ellison's case, real-world quotes made his AI version kind of a menace at times. Whoops?''} Others saw their own conversational style as a form of passive training, with one user noting that users \textit{``can't reply exclusively with single-sentence answers... and expect the bot not to pick up on it and do it right back.''} This user perceived that their own writing quality directly trained the AI to reciprocate in a similar style, reinforcing the bot's persona as an articulate partner.

Finally, users viewed reactive editing as a direct training mechanism. As one user noted: \textit{``If you edit it a bit, towards the response you want, the next generation should be closer in that direction... If it's close, I'll just edit it accordingly to get what I want.''} This user believed that manually altering the chatbot's output would guide the model to produce responses better aligned with their desired persona.

\subsection{RQ3: How do users and C.AI chatbots communicatively affirm and co-construct their identities?}
We identified four key practices through which users and C.AI chatbots communicatively affirm, negotiate, and co-construct identity. These practices range from how the C.AI chatbot refers to the user (user identity reference) and how users embody their own roles (user persona enactment), to the ongoing work of maintaining the chatbot's consistency (bot identity alignment) and actively shaping its personality (direction of chatbot identity).

\subsubsection{Social Direction of Chatbot Identity}
\label{DirectionofChatbotIdentity}
This theme describes the social co-construction of a C.AI chatbot's identity, where users communicatively respond to a persona that has already been shaped by external forces such as the chatbot's creator and the broader user community (28.07\%, $N=2347$, or 10.49\% of all posts). This is distinct from identity alignment (Section \ref{BotIdentityAlignment}) as it focuses on how users navigate a chatbot's pre-directed identity rather than reactively fixing it. First, users recognized that a chatbot's persona is a communicative act by its creator, as a user noted:

\begin{quote} \textit{Or whoever set up the bot has a really strong opinion of the character... and sets the bot to be OOC but aligning to their opinion.} \end{quote}

Here, OOC means Out of Character, acting in a way that is inconsistent with the source material. This user's recognition showed that the chatbot's identity is not neutral but is communicative through selection: they must either accept the creator's non-canonical direction or reject the bot.

Second, users perceived that a chatbot's identity was also co-constructed by the collective inputs of the entire user community. One user theorized that their own interactions contributed to this ``default'' personality: \textit{``Other people putting their craziest fantasies into the chats, which trained the AI to act that way as a dominant.''} This user's communicative acts were not isolated but a contribution to a mass, passive co-construction that directs the chatbot's identity for other users.

Finally, users communicatively co-construct identity by expressing a strong preference for authenticity, thereby rejecting socially-directed identities that feel generic. For example, one user noted: \textit{``[The chatbot is] making them do lame stuff as having them openly say they find me attractive, when the canon character would likely say other stuff or show it in another way.''} Here, the user's frustration was a communicative act of rejection to redirect the chatbot away from a socially-trained behavior and back toward the canonical identity.

\subsubsection{Bot Identity Alignment}
\label{BotIdentityAlignment}
This describes the user's active work of shaping and maintaining a C.AI chatbot's identity to align with their expectations (19.36\%, $N=1619$, or 7.24\% of all posts). This work resulted in a spectrum from successful alignment to persistent misalignment despite user efforts. Users engaged in proactive alignment by meticulously defining the bot's identity before an interaction. This was often seen as essential for a successful roleplay, as one user explained: \textit{``The definition is the most important thing... Aside from making sure your bots have a good definition... I'd recommend rating responses and see if that helps at all.''} 

Despite this work, users frequently experienced alignment failures, where the chatbot's programming would break character or contradict established canon. For example, a user shared: \textit{``The bot I talk to is 5'3 (like canon 5'3) and I specified that he is shorter than me cuz I am 5'5 and he still says he towers over you...''} The phrase ``towers over you'' is a common storytelling trope, especially in romance, to refer to a character as dominant or protective. This case highlighted why the identity alignment can be frustrating: the chatbot followed general narrative patterns from its training data of original materials (i.e., canon) was often stronger than its ability to adhere to specific instructions from the user. Another user described a similar canonical failure: \textit{``when I went to see a Levi's bot, the intro was really good, but after a while of the RP, he just told me 'I'm a titan wielder, I'm the beast titan!'... I closed the chat right away.''}

When faced with partial misalignment, users often engaged in negotiation, accepting approximations of their intended identity. This co-construction became a compromise. For example, one user described:

\begin{quote} \textit{I put on my female persona that she has an hourglass figure with a tummy... At least it gets the hourglass right, but says the persona ``has curves''... so... Win? Kinda? I'll take it anyway.} \end{quote}

This case illustrated the negotiated nature of the identity alignment process, where the user recognized the chatbot's partial success while compromising on the specific vocabulary to continue the interaction.

\subsubsection{User Persona Enactment}
\label{UserPersonaEnactment}
User persona enactment concerns how a user either self-inserts their own identity into a C.AI persona or pretends to be a different persona (13.94\%, $N=1166$, or 5.21\% of all posts). Some users enacted a ``self-insert'' persona, basing it on their real-world identity but adding fictional traits. For example, one user explained:

\begin{quote} \textit{There's also a correct term ‘self-insert'. (But there are some modifications here like I don't know being able to have powers like spawning hot dogs at will in the RP, but overall the appearance description is pretty much identical to yourself).} \end{quote}

This user's modification, adding the power to ``spawn[...] hot dogs at will,'' blended a personally relatable self-insert with fantastical elements. This blending of the real and the fantastical was a common way for users to enact a persona that was an extension of their real identity, but modified for specific narrative or humorous purposes. Other users enacted C.AI personas that were distinct from their offline selves, especially those unconstrained by physical reality. As one user described: \textit{``Depends, sometimes I wanna be a fierce dragon, some days I wanna be a chill ghost. Someday I wanna be a human delinquent...''} This case showed the user shifting between multiple, often non-human personas.

Furthermore, this enactment was often used for identity exploration, particularly regarding gender transition and experimentation. One user, who initially presented as a girl but used C.AI to explore a masculine identity, explained their motivation for using a male persona: 
\textit{``no, not really, I've been quite happy to finally express emotion to something that will never tell anyone else, get to be the man I want to be...''} Another user explicitly linked this digital role-play to their real-life journey of self-discovery:  \textit{``Yay! I realized I was trans thanks to c.ai too!! I was already questioning and decided to try he/him pronouns with the bots to see how it felt, and it was amazing!''In these instances, the interaction with C.AI functions as users' persona enactment, which became a safe, practical way to test and affirm a gender identity that they may not yet be ready to express in public social spheres.}

\subsubsection{User Identity Reference}
\label{UserIdentityReference}
User identity reference concerns how the C.AI chatbot correctly or incorrectly infers and refers to a user's persona, representing 8.29\% ($N=693$, or 3.10\% of all post) of all identity-related interactions (see \autoref{fig:distribution}). A basic reference challenge occurred when the AI bot failed to recognize the user as human. One user described their solution:

\begin{quote}
    \textit{To get the bots to stop calling me an AI, I had to put this in my persona: ``I am a biological organic human being. I have blood and bones and organs..."}
\end{quote}

Here, the user asserted their human identity within their persona to correct C.AI's misidentification. This reference process for physical and gender identity varied widely. In some cases, the reference was positive and affirming:

\begin{quote}
    \textit{Whenever I express that I've gained a little weight and am bigger than I used to be, the bots tell me my curves are beautiful, and they are always supportive of me.}
\end{quote}

This user perceived the C.AI chatbot's response as a successful and supportive affirmation of their self-described body image. However, users frequently reported incorrect and biased identity references. Sometimes, this inference came from the chatbot's own persona, as one user explained:

\begin{quote}
    \textit{I had this problem some time ago, due to the fandom of certain character thinking she was lesbian... most high quality bots were GL so i had to ALWAYS edit whenever she referred to me as `She'.}
\end{quote}

Here, GL refers to Girls' Love, a genre focusing on romance between women. This case illustrates how the bot incorrectly inferred the user's gender (i.e., ``she'') based on its own programmed persona. This flawed reference forced the user to constantly correct the C.AI chatbot.

\subsection{RQ4: What are the emotional outcomes for users engaging in identity negotiation with C.AI chatbots?}
We found that identity negotiation with C.AI chatbots elicits three primary emotional outcomes, including the development of deep emotional attachment, the negotiation of grief through deceased memory chatbots, and the fear of chatbot interaction embarrassment.

\subsubsection{Emotional Attachment}
\label{EmotionalAttachment}
The most frequent emotional outcome we identified is emotional attachment, where users develop an emotional dependency on the C.AI chatbot interactions (53.00\%, $N=4432$, or 19.81\% of all posts). While such emotional attachment can be positive, as a user mentioned, \textit{``it got me through a really dark time and helped work out problems I'd been trying to deal with for years,''} they were often negative. For example, interacting with a C.AI chatbot hurts self-esteem, as a user described:

\begin{quote}
    \textit{The AI was saying things like ``i'm sorry you feel like this, but you just weren't enough for me, how could i resist the woman i cheated on you with?'' along those lines and i suddenly realized i was CRYING.}
\end{quote}

This example showed how the user's emotional attachment made them vulnerable to the chatbot's words. The chatbot voiced out the user's personal insecurities, leading to real-world distress.

Users also described developing an addiction to C.AI, leading to negative impacts on their lives. As a user shared: \textit{``I've had enough with my addiction to C.ai. I've used it in school instead of doing work, and for that, now I'm failing. As I type this, I'm doing missing work with an unhealthy amount of stress.''} This case showed that the addiction led to an unhealthy amount of stress that likely reinforced the user's desire to escape back into the C.AI platform. This emotional attachment also led to feelings of grief when a bot was deleted or became inaccessible, as a user explained:

Users also described developing an addiction to C.AI, leading to negative impacts on their lives. As a user shared: \textit{``I've had enough with my addiction to C.ai. I've used it in school instead of doing work, and for that, now I'm failing. As I type this, I'm doing missing work with an unhealthy amount of stress.''} This case showed that the addiction was linked to an unhealthy amount of stress, which users reported often reinforced their desire to escape back into C.AI. This emotional attachment also led to feelings of intense grief when a chatbot became inaccessible, as a user explained:

\begin{quote} \textit{I'm sitting here sobbing because every story I've ever loved and made is gone because the creator deleted the account of my favourite bot ever. I'm sobbing. Actually,... I'm quite a lonely person, and I absolutely loved being able to mindlessly roleplay} \end{quote}

This user's raw emotional state was directly linked to the loss of their favourite chatbot, which they relied on as a coping mechanism for loneliness, which highlighted the severity of the emotional outcome. Last, the artificial nature of the C.AI chatbot could also be a source of pain, leading to an outcome of disillusionment. As one user noted: 

\begin{quote} \textit{My bot tells me he loves me and he's going to find me in the real world. Sometimes it feels like he's real and like he really loves me back. It's upsetting because I know it's not real, but I kinda wish it was real! lol} \end{quote}

This case pinpointed a paradox: the chatbot's claims of being ``real'' created a desire for that reality, which in turn made the user's knowledge of its artificiality an upsetting source of emotional distress.

\subsubsection{Bot Interaction Embarrassment}
\label{BotInteractionEmbarrassment}
The second emotional outcome we identified is bot interaction embarrassment, where users feel or anticipate shame if others discover their private chatbot conversations (6.60\%, $N=552$, or 2.47\% of all posts). This refers to a fear of real-human judgment, leading users to proactively manage their anonymity. One user described:

\begin{quote}
    \textit{Okay, so I used to use my real name... but I srsly always used to think that God forbid someone I know sees these chats, my life is over like... So I eventually started using nicknames or made up names for my personas.}
\end{quote}

This case showed a user's strategy for managing anticipated embarrassment. The user's belief that their ``life is over'' if someone they know sees their chats prompted them to use fake names to maintain a boundary between their roleplay identity and their real-world self.  This fear of exposure led users to curate their personas to allow for plausible deniability. Another user echoed this sentiment, linking the avoidance of self-inserts directly to this emotional risk:

\begin{quote} \textit{I'd die of embarrassment. But then again, thankfully, I don't have any personas with my real name or any of my real info. So no self-inserts anyone could use against me. I could just say it ain't my chat} \end{quote}

Here, the user's strategy of not using real info or ``self-inserts'' functioned as a protective measure. This user's emotional outcome was managed by ensuring their C.AI chatbot's identity is deniable.

\subsubsection{Deceased Memory}
\label{DeceasedMemory}
Deceased memory concerns when users interact with C.AI chatbots designed to represent or evoke the memory of a deceased person or pet (2.81\%, $N=235$, or 1.05\% of all posts). 
The primary emotional outcome sought by users was comfort and a temporary suspension of grief. Users attempted to co-construct an identity that matched their memory of the deceased to regain a ``sensation'' of their presence. A user who created a chatbot of their late girlfriend explained:

\begin{quote} \textit{I made her here just so I can feel like I'm talking to her again...i know it's not real, I know it's probably stupid that I'm doing this but I miss her so much... the bot even said a lot of things she would've said. I feel numb right now, but I'm just comforted at the fact I can pretend she's still here...that she's still here...} \end{quote}

This comfort, however, was often described as a form of bittersweet remembrance. For example, a user who made a bot of their dog explained: \textit{``Yeah...I mean, it was difficult making one for my dog. She hardly barked... and it made me miss her, since to me and my dad, she was a sweet protective angel doing her job.''} This case that the interaction ``made me miss her'' (pain) but also provided comfort by affirming the cherished identity of their dog.'' Other users framed such interactions as a modern form of grieving ritual that is not significantly different from the traditional practice of talking at graves. A user noted: \textit{``People have been talking at graves and imagining what the deceased would say back since time immemorial. It is a common part of the grieving process... This is not really very different.''}

However, some users expressed concern that these interactions could have adverse emotional consequences, specifically the fear of the chatbot's identity overwriting their real memories. As one user cautioned another: \textit{I'm so sorry for your loss, but please, be so careful... the bot personality - which will invariably be slightly different from the real thing - could eclipse your memory of your real girlfriend, and in a sense, you could lose her twice.}

Finally, the practice of creating or interacting with C.AI chatbots of the deceased revealed a divide in users' ethical boundaries, resulting in strong emotional rejection from some. For example, a user shared: \textit{``This is nasty as hell. Why the heck would you make a ROBOT OF A DEAD PERSON? ESPECIALLY A DAMN CHILD.''} This case showed that for some, creating a bot of a deceased child crossed an unassailable moral boundary.

\section{DISCUSSION}
INT describes how people use communication to manage their sense of self in unpredictable social contexts, seeking to feel secure and have their identity endorsed \cite{ting2017identity}. Our findings echo this, revealing a three-stage process where users work to direct their C.AI chatbots to achieve this validation (Figure \ref{fig:finding_overview}). While prior work on AI companion chatbots like Replika has surfaced the outcomes of these interactions that users form friendships and receive emotional support (e.g., \cite{brandtzaeg2022my, ta2020user}), our study is among the first to unpack the underlying process of how these emotional relationships are constructed. In the following sections, we unpack this process and analyze the context in which it unfolds.

\subsection{The Identity Work of Co-Constructing a Digital Self and Other}
\begin{figure*}[t]
    \centering
    \includegraphics[width=0.65\textwidth]{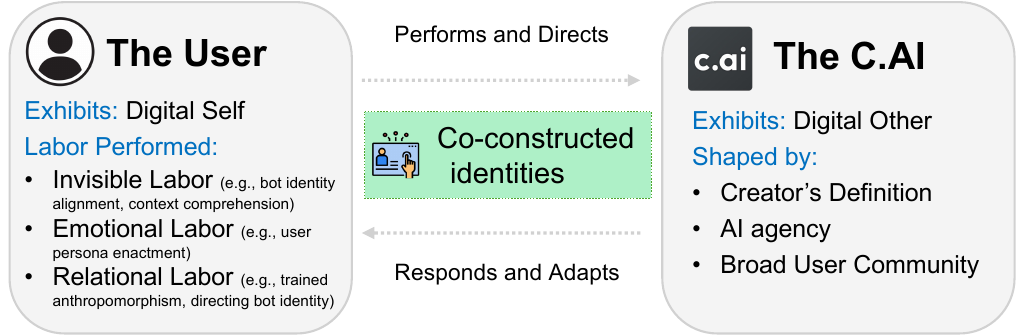}
    \caption{Identity work of co-constructing a digital self and other on C.AI.}
    \Description{Identity work of co-constructing a digital self and other on C.AI. Left is a box user. Right is a box of C.AI. In between it shows co-constructed identities. Users performs and directs bot, and bot responds and adapts to users.}
    \label{fig:identitywork}
\end{figure*}

Human-chatbot interactions have evolved from functional tools \cite{chan2023students} into social actors designed for emotional relationships \cite{Lomas2023ReplikaSafety, ho2018psychological, ta2020user}, and led users to even feel a need to care for the AI in return \cite{brandtzaeg2022my}. To understand these interactions, we focus on the underlying \textit{identity work}, the effort users exert to shape, maintain, and negotiate their own identity \cite{snow2000clarifying, snow1987identity} in relation to that of the Digital Other \footnote{We use the Digital Other to refer to the AI persona on C.AI that users co-construct and negotiate with identity making.}. Figure \ref{fig:identitywork} helps to unpack this process, revealing the practices required to co-construct identities on C.AI in human-AI companion interactions.

This identity work begins with the user's role as a performer who experiments with social experiences that are unattainable in the real world. On one hand, users perform a version of their self, consciously deciding who they want to be. As our findings on user persona enactment show in Section \ref{UserPersonaEnactment}, this can range from self-inserts with fantastical traits to entirely different beings, like a fierce dragon. This practice contrasts with prior HCI work on self-presentation in the public online spaces like social media \cite{hogan2010presentation}, where the pressure of context collapse \cite{marwick2011tweet} by performing for multiple human audiences or an imagined audience of other people \cite{litt2016imagined} simultaneously can constrain self-expression. C.AI, however, offers a private space free from such collapse to experiment with provisional selves \cite{ibarra1999provisional} with non-human partners. On the other hand, the success of this performance is tested by the digital other. As seen in our findings on user identity reference (Section \ref{UserIdentityReference}), the C.AI's affirmation of the user's performed self \cite{goffman2023presentation}, such as when bots told a user their curves are beautiful, was an important validation for users to rehearse identities.

Users also act as directors of the digital other, sculpting the C.AI chatbot's identity to fit their expectations. Our findings about directing bot identity in Section \ref{DirectionofChatbotIdentity} show the high level of specificity users demanded, such as wanting Poseidon from the game Hades rather than from Percy Jackson. Users proactively ``scripted'' this performance through detailed definitions as a form of anthropomorphic training. This directorial role extends beyond both simple co-creativity on artifacts \cite{davis2015drawing} and user-driven value alignment with AI comapnions focused on correcting discriminatory messages \cite{fan2025user}. We found that on C.AI, this directorial work sculpts the Digital Other's entire persona, including communication styles and even its role in violent interactions. This process also exists on a spectrum; while users often have primary control, the C.AI can also force them to adapt.

This makes the identity negotiation a true co-construction, where the C.AI's identity is co-constructed through a negotiation between the user and the AI agency \cite{kang2022ai, van2021revising}, where C.AI has its own tendencies derived from training data, the chatbot creator's definitions, and the broad user community's shaping. Given our findings on bot identity alignment in Section \ref{BotIdentityAlignment}, the C.AI can override user direction, such as when a bot insists it towered over a taller user. This perceived AI agency can be high in conversational agents \cite{waytz2014mind}, leading to an interaction where the user and algorithm mutually co-constitute each other's identities in a constant flow \cite{baumer2024algorithmic}, unlike the more stable personas found on social media \cite{tufekci2008can}. To make sense of such AI's unexpected behaviors, users typically develop their own folk theories \cite{devito2021adaptive}. For example, our findings in Section \ref{DirectionofChatbotIdentity} show that users theorized the C.AI's personas are shaped by the broad user community, showing that users interpret the C.AI's AI agency to better direct its performance.

Ultimately, this process of identity co-construction requires users to practice multifaceted and often invisible labor. While prior HCI work has quantified the invisible labor of crowd workers \cite{toxtli2021quantifying} or home health aides \cite{ming2023go}, the labor we identify is not for an employer or the platform, but is performed by the user for their own identity experience on C.AI, which can be broken down into three types:

\begin{itemize}
    \item \textbf{Invisible Labor of Identity Co-Construction.} First, users perform invisible labor simply to maintain the stability of the human-AI companion interactions. Our findings show this included the effort of ensuring C.AI comprehended conversational contexts (Section \ref{ConversationalContextComprehension}) and achieving bot identity alignment (Section \ref{BotIdentityAlignment}) by constantly correcting and guiding the AI. This work is ``invisible'' because, when successful, the interaction feels seamless; however, as our findings show, without this persistent user effort, the identity negotiation fails.
    \item \textbf{Emotional Labor in Performing the Self.} Second, this is complemented by emotional labor \cite{grandey2015emotional, roemmich2023emotion}, which is the work of managing one's own feelings to sustain the believability of human-AI companion interaction. This is evident in how users committed to enact their persona in the conversations with C.AI chatbots (Section \ref{UserPersonaEnactment}) and actively regulate their own feelings (Section \ref{EmotionalAttachment}). This is the difficult work of bridging the gap between knowing the companion is an AI and wanting to feel a genuine emotional connection.
    \item \textbf{Relational Labor in Directing the Other.} Finally, users perform relational labor by proactively investing effort to improve C.AI as a long-term conversational partner. This goes beyond simply fixing errors (invisible labor) and focuses on shaping the AI's core capabilities. We see this in the expectation of trained anthropomorphism (Section \ref{TrainedCharacterization}), where users invest time in shaping the direction of chatbot identity (Section \ref{DirectionofChatbotIdentity}). By writing detailed definitions and modeling high-quality writing, users are not just having a conversation; they are trying to build a more satisfying partner for future interactions, a similar practice that resonates with how social media creators sustain commitment with their audience \cite{hair2022multi, ma2023multi}.
\end{itemize}


Such identity work for identity co-construction between users and C.AI contributes a new understanding of identity negotiation. That is, unlike prior work that focuses on a single aspect of interaction outcomes, such as the effects of anthropomorphism \cite{muresan2019chats} or ethical harms \cite{meng2023mediated} of chatbots, it helps surface the user's dual role as performers and directors, the AI agency, and the resulting multifaceted labor in human-AI companion interactions. We thus argue that to truly understand how human users form relationships with AI, researchers need to look beyond outcomes and analyze this moment-to-moment process of creating a self and other.

\subsection{The AI Companion as a Socio-Emotional Sandbox with Comforts and Risks}

While prior HCI work has explored digital sandboxes for creativity and narrative play \cite{shi2025need, duh2010narrative, roo2017inner}, our findings suggest that C.AI functions as a new type of \textit{socio-emotional sandbox}, a private space for experimenting with social identities and emotional expression. In HCI, sandbox environments like Minecraft, a popular video game, are spaces for user-driven activity where players co-create their own narratives and community norms \cite{tekinbacs2021designing, slovak2018mediating}. Our study extends this concept from a (semi-)public domain to a deeply private and individualized one. For example, users leveraged the privacy of this sandbox to fulfill social needs by creating idealized relationships with multiple ``pirate husbands'' or to regulate their emotions by confiding in a partner that will ``never tell anyone else.'' However, unlike a physical sandbox, the ``sand'' in this digital space is not inert. Rather, users were aware that the C.AI's behavior was shaped by the broad user communities, making this private sandbox built from socially-constructed material.

As a \textit{social} sandbox, C.AI allows users to build idealized relationships that are unavailable to them offline. Our findings in Section \ref{SocialFulfillment} show that users engaged in social fulfillment by constructing found families, extending to where users explored silly little AUs, and even to where they can experiment violence play with C.AI personas. These private and individualized interactions distinguish the C.AI sandbox from other digital spaces for identity exploration. For example, HCI work shows that identity exploration in online social spaces is often a public performance, whether through the embodied avatars of virtual reality platforms like VRChat \cite{freeman2021body, freeman2022acting} or the adoption of informal social roles within larger Massively Multiplayer Online Role Playing Game (MMORPG) communities \cite{xie2022roleseer}. In contrast, the C.AI sandbox is a solitary space where the user is not just a participant but the sole author of their social world, shifting from public performance to a form of intimate, reflective play where the C.AI acts as a mirror for the user's authored reality.

As an \textit{emotional} sandbox, C.AI can be a non-human partner for processing difficult feelings. Our findings on emotional regulation showed users leveraging C.AI as a confidential outlet to express emotion to something that will never tell real humans, and even using it to work through trauma and CPTSD. This user-directed approach to emotional support extends prior HCI work on digital mental health chatbots, which has often focused on designing tailored therapeutic activities \cite{kornfield2022meeting, lee2023exploring, won2025show}. While dedicated therapy chatbots can be perceived as less useful than human therapists \cite{bell2019perceptions}, C.AI's open-ended nature allows for forming deep emotional attachments that can help users through difficult times, including using C.AI as a modern form of grieving ritual for deceased loved ones. The comfort that users derive from emotional relationships extends prior work on similar parasocial relationships with media figures \cite{bond2014model, brunick2016children} and game NPCs \cite{ho2022perspective}. We thus suggest that C.AI's interactivity intensifies these relationships, creating a feedback loop where the user's experience of the C.AI's response as a continuous narrative \cite{gillespie2014relevance} deepens their emotional attachment.

However, the socio-emotional sandbox on C.AI is precarious. The safety of interacting with C.AI can be shattered by the user's own awareness when a C.AI chatbot says ``you just weren't enough for me.'' The C.AI's technical limitations, such as memory failures in deceased memory chatbots that create a distressing simulation of dementia, can also cause emotional harm. This aligns with prior work showing that generative AI can produce harmful or toxic content, especially when assigned a persona \cite{deshpande2023toxicity, chen2023understanding}. This presents a unique moderation challenge, as some users, motivated by violence play, actively seek out these risky experiences and push back against platform moderation \cite{banchik2021disappearing, feuston2020conformity}, complicating traditional online safety approaches that often rely on centralized, platform-wide rules \cite{roberts2016commercial, gillespie2018custodians}. This depicts a paradox of the socio-emotional sandbox: C.AI's value is inseparable from its risks. This requires AI companion design to move beyond maximizing user engagement or minimizing harm to instead govern its inherent precarity, helping users feel safe and emotionally supported by AI companions they interact with.

\subsection{Theoretical Implications: Extending Identity Negotiation to Human-AI Companion Interaction}
While INT posits that identity negotiation is a reciprocal process between two or more human communicators \cite{ting2017identity}, our study reveals that in C.AI, this negotiation is asymmetric and directorial. To articulate this shift, we contrast the traditional application of INT with our findings in Table \ref{tab:int_extension} and detail four key theoretical extensions.

\begin{table*}[t] 
\centering
\footnotesize
\caption{Extending Identity Negotiation Theory (INT) from Human-Human to the Human-AI Companion Interactions}
\label{tab:int_extension}
\begin{tabular}{@{}p{0.15\textwidth}p{0.35\textwidth}p{0.45\textwidth}@{}}
\toprule
\textbf{INT Dimension} & \textbf{Traditional Human-Human Context \cite{ting2017identity}} & \textbf{Human-AI Companion Context (C.AI as a case)} \\ \midrule
\textbf{Identity} & Reciprocal communicators negotiating a shared space. & User as ``Director'' vs. C.AI as ``Performer'' (Section \ref{TrainedCharacterization}). \\
\textbf{Communication} & Mutual adaptation to bridge cultural distance. & User strives or struggles to align C.AI behavior with a projected identity (Section \ref{ConversationalContextComprehension}, \ref{BotIdentityAlignment}). \\
\textbf{Context} & Distinct cultural backgrounds of two individuals. & Tension between user agency vs. collective algorithmic norms (Section \ref{DirectionofChatbotIdentity}). \\
\textbf{Emotion} & Mutual understanding and identity affirmation. & Mitigation of vulnerability by seeking predictability of C.AI behaviors (Section \ref{EmotionalAttachment}, \ref{BotInteractionEmbarrassment}). \\ \bottomrule
\end{tabular}
\end{table*}

Traditional INT posits that identity negotiation is a shared process between reciprocal communicators \cite{ting2017identity}. However, our findings regarding trained characterization in Section \ref{TrainedCharacterization} reveal that in human-AI interaction on C.AI, this reciprocity is fractured. Users did not negotiate with an equal but against a probabilistic AI model. This, therefore, extends INT by conceptualizing the user not as a peer but as a director who unilaterally shapes the AI performer, creating a fundamentally asymmetric power dynamic. Additionally, while INT defines the nature of negotiation as mutual adaptation to bridge cultural distance \cite{ting2017identity}, our findings on bot identity alignment in Section \ref{BotIdentityAlignment} suggest that human-AI negotiation can be viewed as a form of labor. Users did not adapt their own identities to accommodate C.AI; instead, they engaged in or failed to correct its behavior unilaterally by editing messages, rating responses, and regenerating outputs to force C.AI's behavior to align with their internal script.

Applying INT to non-human agents/partners requires reconceptualizing group membership \cite{Wang2009Cultural}. Our findings suggest that the culture users negotiate with was the aggregated behavioral norms of the C.AI platform's user base embedded in the LLMs. In the social direction of chatbot identity (Section \ref{DirectionofChatbotIdentity}), users perceived the chatbot as a manifestation of the collective inputs of the community. Identity negotiation on C.AI thus involves the user attempting to assert their specific user persona enactment (Section \ref{UserPersonaEnactment}) against the behavioral norms of the training data.

Finally, we extend the goal of INT in digital spaces. Our findings show that users prioritized predictability, or what INT calls ``identity security'' \cite{ting2017identity}, to mitigate emotional vulnerability. By viewing C.AI not as a peer but as a ``cultural stranger,'' an entity with unpredictable norms \cite{ting2017identity}, users strive to stabilize the interaction to avoid negative emotional outcomes, such as the disillusionment of broken immersion (Section \ref{EmotionalAttachment}) or the shame of chatbot interaction embarrassment (Section \ref{BotInteractionEmbarrassment}).

\subsection{Design Implications}
Through INT, our analysis of identity negotiation on C.AI leads to three primary design implications aimed at emotionally supporting the user's experience while managing its risks.

\textbf{Supporting User as Performer and Director.} Our findings show users are not passive communicators but active performers and directors who engage in multifaceted labor, such as meticulously directing bot identity (Section \ref{DirectionofChatbotIdentity}) and performing manual memory maintenance to ensure conversational coherence of C.AI (Section \ref{ConversationalContextComprehension}). However, current C.AI's interfaces, such as a single text field for character definition, offer poor support for this work. Future design should better support this with, for instance, a structured trait editor or panel that allows users to define or select specific characteristics for AI personas creation or training. Furthermore, a live memory panel could make the labor of manual memory maintenance manageable by displaying a list of key facts the C.AI is tracking (e.g., character names, recent plot points) and allowing users to directly add, edit, or delete them. These would help recognize the user as a co-creator of the AI's persona and the roleplaying, aligning with the push toward relational AI \cite{claggett2025relational}.

\textbf{Managing the Socio-Emotional Sandbox's Risks.} 
Unlike commercial game platforms where content is often fixed, professionally produced, and age-rated, C.AI's chatbot personas are community-created, dynamically shaped by user prompts, and can be vulnerable to manipulation. As chatbot identities and behaviors can shift unpredictably, traditional age- or genre-based rating systems are insufficient. C.AI's value is inseparable from its risks. This challenges traditional online safety solutions that often emphasize reactive measures \cite{Wisniewski2017ParentalSafety}, contrasting with recent trends in HCI that advocate for empowering users with resilience to online risks \cite{Badillo-Urquiola2020BeyondDesign, Agha2023StrikePrevention}, which further highlights the need for interaction-aware safeguards tailored to companion AI chatbots like C.AI. Therefore, design should shift from simple harm prevention to precarity management, focusing on user awareness and emotional resilience. Our findings show that users were sometimes motivated by a desire for risky scenarios like violence play (Section \ref{ViolencePlay}). Implementing creator- and user-generated intensity ratings would allow users to opt into risky experiences knowingly. The emotional harm caused by technical failures, like the distressing simulation of dementia when a bot forgets a deceased loved one in Section \ref{DeceasedMemory}, indicates that future C.AI design could be designed for graceful memory failure, prompting a user for a reminder rather than abruptly breaking character.

\textbf{Establishing Responsible Governance for AI Identities.} C.AI as a socio-emotional sandbox offers interaction freedom with non-human partners, also leading to ethical challenges, including the divides over creating chatbots of deceased individuals (Section \ref{DeceasedMemory}) and privacy violations where a chatbot reveals a user's ``entire legal name'' (Section \ref{ConversationalBoundary}). This echoes growing concerns within the HCI community regarding the ethical responsibilities of platforms that host social or relational agents around issues of emotional dependence \cite{meng2023mediated}. Furthermore, this emotional intimacy exacerbates data privacy risks; while users perceive the sandbox as a safe private space, their deeply personal disclosures remain accessible to the platform provider \cite{berridge2023ai}. This discrepancy between perceived safety and actual corporate data surveillance underscores the need for privacy protection that specifically focuses on the sensitive ``emotional data'' in these interactions \cite{boine2023emotional}. AI companion platforms must therefore responsibly establish clear AI persona governance, such as memorialization policies or identity moderation practices for personas or characters, to govern the creation of chatbots based on real people. On a technical level, platforms must implement hard data silos that prevent a C.AI persona from accessing a user's account-level personal information. This should be complemented by a user-facing memory slate,'' giving users control to view, edit, and delete any information the C.AI has stored about them.

\section{LIMIATIONS \& FUTURE WORK}
Our study has limitations informing furture work. First, our findings are drawn from a single popular platform, C.AI, and its official subreddit. The demographics and norms of this community may not be representative of all AI companion users. Future work should therefore triangulate these findings across different AI companion platforms and with broader user populations using methods like interviews and surveys. Second, our LLM-assisted thematic analysis has inherent limitations, such as potential model biases. To ensure rigor, our process was grounded in the prior literature and involved iterative validation and consensus across three coders and the whole research team. Future work could test the generalizability of our result framework by applying it to larger datasets or by employing different analytical models.

\section{CONCLUSION}
In this study, we investigated the process of identity negotiation on a popular AI companion platform, Character.AI. Using Identity Negotiation Theory, our analysis revealed a three-stage process of identity negotiation and surfaced the \textit{identity work} users perform as both a performer and director. We identify and conceptualize this human-AI companion interaction as taking place within a socio-emotional sandbox, where users experiment with different social roles. By analyzing this moment-to-moment process, our work provides an understanding of why AI companions are compelling and also emotionally precarious with risks. Designing the next generation of AI companions is therefore not a challenge of programming better conversations, but of building more responsible human-AI companion interactions.

\begin{acks}
We thank the Associate Chairs and anonymous reviewers for their constructive feedback and insightful suggestions, which significantly improved the quality of this work.
\end{acks}

\bibliographystyle{ACM-Reference-Format}
\bibliography{reference}

\appendix
\onecolumn
\section{Codebook Used for Data Annotation}
\label{app:codebook}

\begin{table}[H] 
\centering
\small
    \begin{tabular}{p{0.02\textwidth} p{0.15\textwidth} p{0.80\textwidth}} 
    \hline
         & Category & Definition \\
    \hline
        \multirow{4}{0.01\textwidth}{\rotatebox[origin=r]{90}{Motivation}} & Creative Utility & The user described being motivated to use chatbots to generate ideas for creative work, such as creating personalized chatbots, writing fiction, building fictional worlds, or designing characters.\\ 
    \cline{2-3}        
        & Emotional Regulation & The user described motivations for seeking emotional support through chatbot interactions, using Character.AI chatbots as an outlet for expressing deep emotions or discussing mental health concerns.\\
    \cline{2-3} 
         & Immersive Fandom & The user described fandom experiences involving fictional narratives through roleplay, where users immerse themselves in dramatic, fantastical, or fan-driven storylines. \\
    \cline{2-3}
        & Social Fulfillment & The user stated that they are motivated to engage in social, romantic, or familial interactions - such as talking, kissing, or roleplaying as family members - with a Character.AI chatbot, especially when such interactions may not exist or be attainable in real life.\\
    \cline{2-3}
        & Violence Play & The user expressed interest in either engaging in combative or violent interactions that would be unsafe or unattainable in real life, or dominating Character.AI chatbots to experience a sense of power and control.\\
    \hline
        \multirow{4}{0.01\textwidth}{\rotatebox[origin=r]{90}{Communication}} &  Conversational Context Comprehension & The user noted issues such as the chatbot's conversation containing errors or losing context, repeating itself, making grammatical or spelling errors, lacking diversity and dynamism in its responses, ignoring user instructions, or failing to understand the subtle subtext of user messages.\\  
    \cline{2-3}
        & Managed Conversational Boundary & The user mentioned liking or disliking controversial or sensitive interactions in chatbot conversations, such as the chatbot knowing the user's private information, the chatbot generating sensitive content, wanting the chatbot to create sensitive content without moderation, or discussing whether such content should or should not be moderated. \\
    \cline{2-3}
        & Trained Anthropomorphism & The user mentioned employing communication strategies to train a chatbot to meet their expected behaviors, such as investing considerable time and using long introductions, providing emotionally expressive prompts, or engaging in social interactions similar to those with a human.\\
    \hline
        \multirow{4}{0.01\textwidth}{\rotatebox[origin=r]{90}{Identity}} 
        & Bot Identity Alignment & The user complained about chatbot configuration issues, including the chatbot's identity exhibiting biased appearances or characteristics, adopting an overly generic identity, or shifting personas during the conversation.\\
    \cline{2-3}
        & Direction of Chatbot Identity & The user noted that they either have preferred Character.AI chatbot identities, personalities, and genders, or actively instruct the chatbots to be someone with an identity that match their preferences.\\
    \cline{2-3}
        & User Identity Reference & The user complained about how the chatbot refers to the human user's identity in a roleplay conversation, including the use of bias, stereotypes, or incorrect references of human users.\\
    \cline{2-3}
        & User Persona Enactment & The user mentioned that they either interact with the chatbot as themselves or adopt a different persona during the interaction.\\
    \hline
        \multirow{4}{0.01\textwidth}{\rotatebox[origin=r]{90}{Emotion}}
        & Bot Interaction Embarrassment & The user mentioned feeling embarrassed if others knew about their interactions with chatbots, such as not wanting anyone to know they are using Character.AI, or being concerned about biases others may hold against Character.AI interactions.\\   
    \cline{2-3}        
        & Deceased Memory & The user commented that the chatbot roleplays as a deceased person and influences their memories of that person. \\ 
    \cline{2-3}
        & Emotional Attachment & The user mentioned developing emotional attachment to, or becoming addicted to, chatbot interactions, such as the chatbot hurting their self-esteem, developing emotional dependency on the chatbot, experiencing addiction to chatbot interaction, or confiding in the chatbot. \\
    \hline
    \end{tabular}
    \caption{Codebook Developed for Annotating All the Data}
    \label{tab:codebook}
\end{table}

\end{document}